\documentclass[12pt]{article}
\usepackage{epsfig}

\def\beq{\begin{equation}}
\def\eeq#1{\label{#1}\end{equation}}
\def\eeqn{\end{equation}}

\def\beqa{\begin{eqnarray}}
\def\eeqa#1{\label{#1}\end{eqnarray}}
\def\eeqan{\end{eqnarray}}

\let\bar=\overbar

\def\Dslash{\not{\hbox{\kern-4pt $D$}}}
\def\dslash{\not{\hbox{\kern-2pt $\del$}}}

\def\msb{{\bar{\ssstyle M \kern -1pt S}}}

\usepackage{fancyhdr,graphicx}
\def\Title#1{\begin{center} {\Large {\bf #1} } \end{center}}

\begin{document}

\Title{The timing behaviour of radio pulsars}

\bigskip\bigskip


\begin{raggedright}

{\it George Hobbs\index{Hobbs, G.}\\
Australia Telescope National Facility\\
CSIRO\\
P.O. Box 76\
Epping, NSW\\
Australia, 1710\\
{\tt Email: george.hobbs@csiro.au}}
\bigskip\bigskip
\end{raggedright}

\section{Introduction}

The remarkable clock-like rotational stability of pulsars has been known since their discovery (Hewish et al. 1968) and has allowed every rotation, over more than 40 years, of PSR~B1919+21 (the first pulsar to be discovered) to be accounted for.  This has been achieved using a technique known as ``pulsar timing'' (an overview is provided in Lorimer \& Kramer 2005) in which software (e.g. \textsc{tempo2}; Hobbs, Edwards \& Manchester 2006) is used to obtain a model for the rotational and positional properties of the pulsar.  The purpose of this paper is 1) to summarise the pulsar timing method, 2) to provide an overview of recent research into the spin-down of pulsars over decadal timescales and 3) to highlight the science that can be achieved using high-precision timing of millisecond pulsars.  Pulsar timing provides input into studies of compact stars by providing accurate estimates of pulsar masses and spin rates as well as allowing the interior of pulsars to be probed through glitch events and studies of long-term spin-down.

Many observatories have pulsar timing projects.  This paper will describe data obtained using the 64-m Parkes radio telescope in Australia and the 76-m Lovell radio telescope at Jodrell Bank, UK.  The Parkes telescope has numerous ongoing pulsar timing programs including 1) the Parkes Pulsar Timing Array project in which 20 millisecond pulsars are observed with the aim of detecting gravitational wave signals (e.g. Hobbs et al. 2009a and references therein), 2) the Fermi project for which $\sim$100 pulsars are monitored in order to provide accurate rotational ephemerides for analysis of gamma-ray data (Weltevrede et al. 2009) and 3) the PULSE@Parkes project in which high-school students carry out monthly monitoring observations of 42 pulsars (Hobbs et al. 2009b).  The Jodrell Bank timing program has mainly concentrated on observing almost all of the known radio pulsars that can be detected from the observatory.  In contrast to the high-precision timing experiments that often use coherent de-dispersion systems and long observing durations, the huge number of pulsars observed at Jodrell Bank Observatory has necessitated short observations.  The Jodrell Bank data archive now contains over 6000 years of pulsar rotational history and therefore provides an ideal data set for studying the long-term spin down of pulsars (Hobbs et al. 2004).

\section{Pulsar timing}

The pulsar timing method relies on multiple measurements of pulse times of arrival (ToAs)\footnote{Typically pulsar signals are too weak for individual pulses to be seen.  Usually hundreds or thousands of pulses are averaged together in order to increase the signal-to-noise ratio and provide a more stable pulse profile.} that have been obtained over a span of days to years at a radio observatory.  From these ToAs, the times of arrival of the pulses as determined by a fictitious clock at the Solar System barycentre are determined. These barycentric arrival times are compared with predicted arrival times obtained from a simple model of the pulsar.  Such models typically contain the pulsar pulse rate and its first derivative, astrometric parameters (position, proper motion and parallax) and any orbital parameters (usually only the Keplerian parameters, but some post-Keplerian parameters may also be required).  The deviations between the predicted and the observed TOAs are known as the pulsar Ôtiming residualsÕ and indicate unmodelled effects or an inaccurate model of the pulsar.  For instance, incorrect determination of the pulsar pulse frequency will lead to the timing residuals having a linear form.  The gradient of the residuals, determined using least-squares-fitting, is used to obtain a better estimate of the pulsar pulse frequency.  This estimate can subsequently be used to produce new timing residuals and the procedure iterated until all the timing residuals are statistically consistent with zero.  However,  timing models currently do not contain any information about all the phenomenon that may effect the measured ToAs.  For instance, strong gravitational wave sources will induce timing residuals (e.g. Detweiler 1974) that, in most cases, will not be removed using the standard pulsar timing fitting process.

Analyses of pulsar timing residuals have implications for a diverse range of astrophysics.  For instance, the first extra-solar planets were detected around PSR~B1257+12 (Wolszczan \& Frail 1992),  evidence of gravitational wave emission was obtained by timing PSR~B1913+16 (Hulse \& Taylor 1975) and the most stringent test of general relativity in the strong-field limit has been provided by PSR~J0737$-$3039A/B (Kramer et al. 2006).  Pulsar timing has also been used to study the interstellar medium (You et al. 2007), to rule out some models of cosmic strings (Jenet et al. 2006), to determine pulsar velocities (e.g. Hobbs et al. 2005) and  to study globular clusters (Freire et al. 2001).

\section{Pulsar timing residuals on decadal timescales}

Examples of the timing residuals observed over decadal timescales for four pulsars are shown in Figure~1 (data from Hobbs, Lyne, Kramer, in press; hereafter Paper~I).  Out of a sample of 366 pulsars (Paper~I),  $\sim$37\% of the pulsars have timing residuals dominated by the measurement errors, 20\% have residuals that take the form of a cubic polynomial corresponding to a positive second derivative of the pulse frequency ($\ddot{\nu} > 0$), 16\% with $\ddot{\nu} < 0$ and the remaining $\sim$27\% have more complex structures. Previously it had been thought that timing residuals exhibited two main types of irregularity: `glitches' in which the pulsar's rotation rate suddenly increases before undergoing a period of relaxation and `timing noise' which consists of low-frequency structures.  

\begin{figure*}
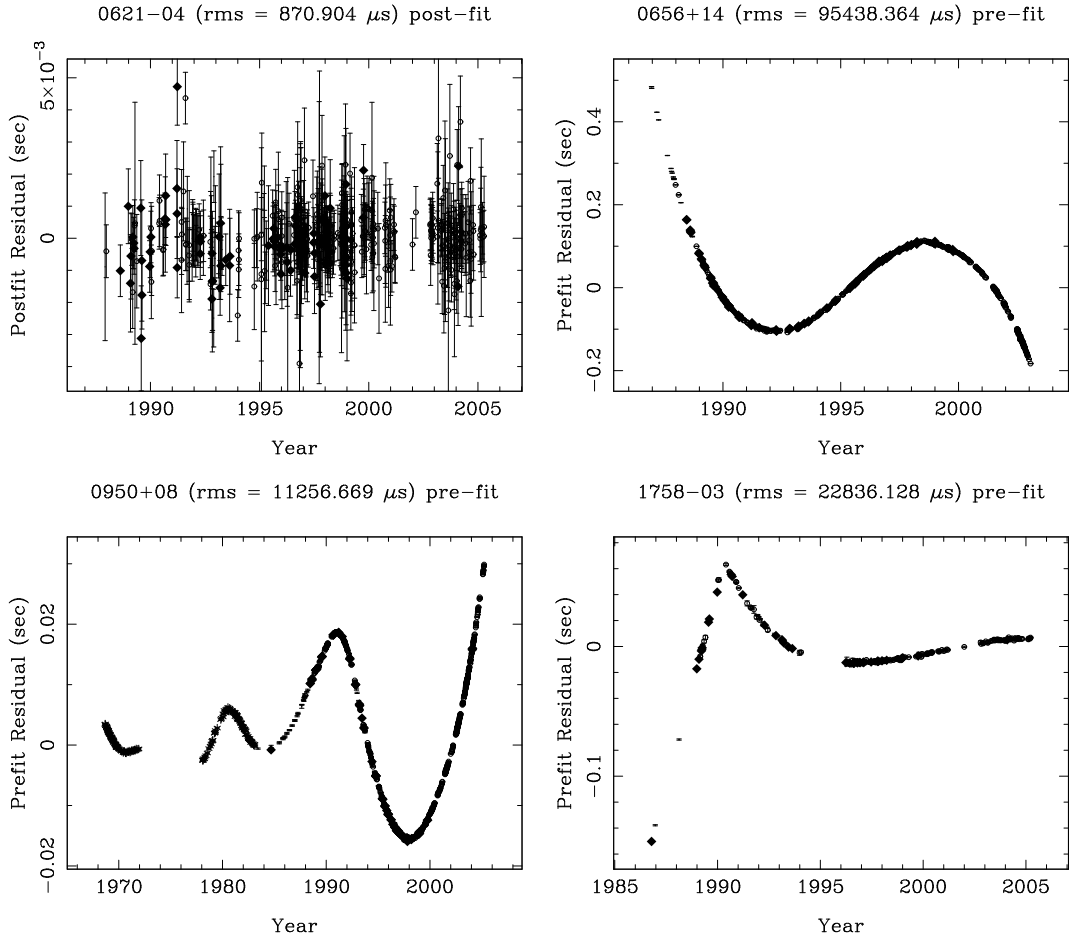

\begin{tabular}{ll}
\includegraphics[width=6cm,angle=-90]{0621.ps} & 
\includegraphics[width=6cm,angle=-90]{0656.ps} \\
\includegraphics[width=6cm,angle=-90]{0950.ps} & 
\includegraphics[width=6cm,angle=-90]{1758.ps} \\
\end{tabular}
\caption{The timing residuals obtained from the Jodrell Bank data archive for PSRs~B0621$-$05, B0656$+$14, B0950$+$08 and B1758$-$03.}\label{fg:timingex}
\end{figure*}

Glitch events can be explained as the sudden unpinning of superfluid vortices in the interior of the neutron star (Alpar et al. 1986).  There is no obvious relationship between timing noise and glitches although Janssen \& Stappers (2006) showed that some timing noise can be modelled using multiple small glitches and Shabanova (1998) introduced the idea of `slow glitches' that can mimic timing noise (both these ideas are argued against in Paper~I).  Melatos, Peralta \& Wyithe (2008) showed that the waiting time between glitch events for many pulsars follows an exponential distribution.  This can be modelled using a cellular automaton model of pulsar glitches (Warszawski \& Melatos 2008).  However, for two pulsars with seemingly similar properties one may glitch regularly and while the other has never been observed to glitch.

The following explanations/descriptions for timing noise exist in the literature:
\begin{itemize}
\item{Random walks in the pulse frequency and/or its derivatives (Boynton et al. 1972)}
\item{Free-precession of the neutron star (Stairs et al. 2000)}
\item{Unmodelled planetary companions (Cordes 1993)}
\item{Magnetospheric effects (Cheng 1987)}
\item{Interstellar/interplanetary medium effects (Scherer et al. 1997)}
\item{Accretion onto the pulsar's surface (Qiao et al. 2003)}
\item{Small glitch events (Janssen \& Stappers 2006)}
\end{itemize} 

\begin{figure*}
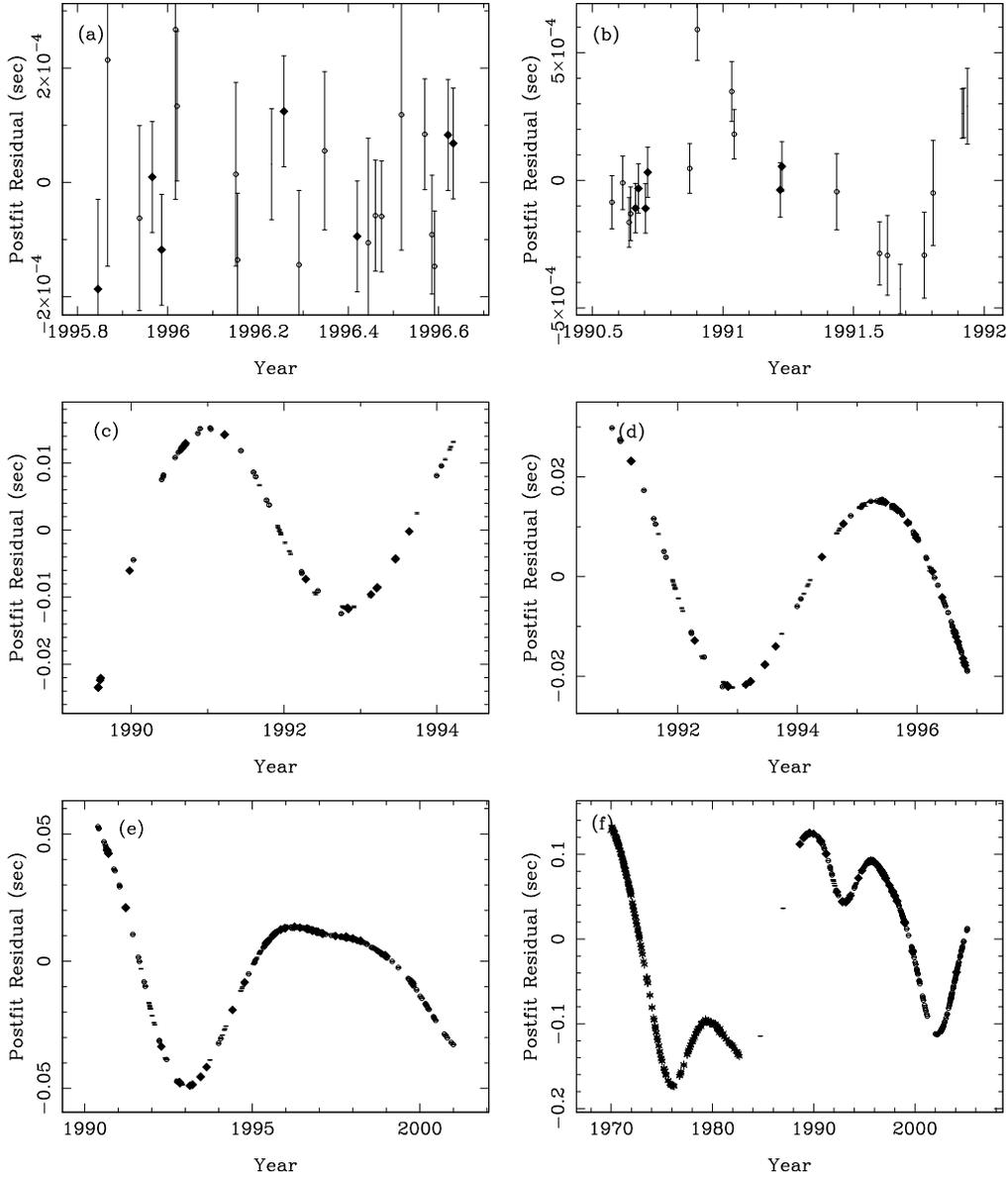

\begin{tabular}{ll}
\includegraphics[width=5cm,angle=-90]{oneY_1.ps} &
\includegraphics[width=5cm,angle=-90]{1.5_yr.ps} \\
\includegraphics[width=5cm,angle=-90]{fiveY.ps} & 
\includegraphics[width=5cm,angle=-90]{sixY.ps} \\
\includegraphics[width=5cm,angle=-90]{elevenY.ps} & 
\includegraphics[width=5cm,angle=-90]{full.ps} \\
\end{tabular}
\caption{The timing residuals for PSR~B1818$-$04 obtained from different section of the entire data-span available.  In each section the pulsar's spin-frequency and its first derivative have been fitted.  The data spans are approximately a) 1\,yr, b) 1.5\,yr, c) 5\,yr, d) 6\,yr, e) 11\,yr and f) the full 35\,yr. Figure taken from Hobbs, Lyne \& Kramer (in press).}\label{fg:moredata}
\end{figure*}

Paper~I  showed (or confirmed) that timing noise is not caused by the 1) on-line data processing being carried out at a particular observatory, 2) off-line data processing software, 3) clock corrections required in converting from observatory time standards, 4) use of a Solar System ephemeris, 5) incorrect calibration of the pulse profiles or 6) interstellar/interplanetary medium.  The paper also highlighted the necessity of long data spans in any analysis of pulsar timing noise.   For instance, the structures seen in the timing residuals for any particular pulsar changes as more data are obtained.  In Figure~\ref{fg:moredata} we reproduce the plot given in Paper~I for the timing residuals for PSR~B1818$-$04 obtained with data spanning between 1 and 35 years.  Only with the longest data span does the quasi-periodic nature of the timing noise become clear.

\begin{figure}
\begin{center}
\includegraphics[angle=-90,width=8cm]{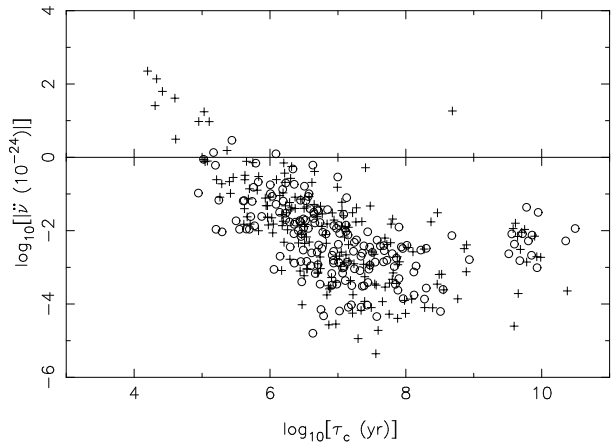}
\end{center}
\caption{The measured $\ddot{\nu}$ values versus characteristic ages. Pulsars with $\ddot{\nu}>0$ are indicated using `plus' signs, with `circles' for those with $\ddot{\nu}<0$.  Figure from Hobbs, Lyne \& Kramer (in press).}\label{fg:nuddot_age}
\end{figure}

One measure of the ``amount of timing noise'' is simply the value obtained by fitting for the second derivative of the spin frequency ($\ddot{\nu}$).  This value is plotted versus characteristic age for the pulsars analysed in Paper~I in Figure~\ref{fg:nuddot_age}.  `Plus' signs in this figure represent values of $\ddot{\nu} > 0$ and `circles' for $\ddot{\nu} < 0$.  The data are consistent with the timing noise for the very youngest pulsars being dominated by glitch recovery ($\ddot{\nu} > 0$; Lyne et al. 2000) whereas the timing noise for older pulsars is caused by a different phenomenon.   

As data spans increase, fewer `cubic' polynomial structures are observed in the timing residuals (after only fitting for the spin frequency and its derivative); the structures become more quasi-periodic in form. This therefore suggests that all current models of timing noise, that are based on noise-like processes, are inconsistent with observations over long time scales.

\section{High-precision millisecond pulsar timing}

\begin{figure}
\begin{center}
\includegraphics[angle=-90,width=8cm]{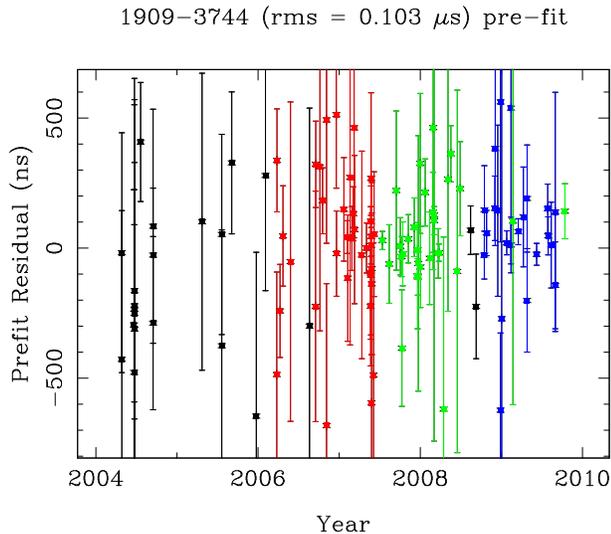}
\end{center}
\caption{Timing residuals at an observing frequency $\sim$3GHz for PSR~J1909$-$3744. The different symbols indicate different backend instrumentation.}\label{fg:1909}
\end{figure}
Because of their fast spin rates and intrinsic stability, the rms timing residuals for millisecond pulsars are typically much lower than for normal pulsars.  The Parkes Pulsar Timing Array (PPTA; Manchester 2008) is one of a few projects worldwide (e.g. Hobbs et al. 2009c) attempting to obtain high quality data sets on $\sim$20 pulsars in order to 1) detect gravitational waves, 2) improve the solar-system planetary ephemeris and 3) search for irregularities in terrestrial time standards.

Theoretical calculations (Jenet et al. 2005) show that, in order to achieve our main goal of gravitational wave detection, $\sim$20 pulsars need to be observed over a five year dataspan with rms timing residuals   $\sim$100\,ns (without fitting for the second derivative, or higher derivatives, of the pulsar's spin frequency).  You et al. (2007) showed that dispersion measure variations caused by the interstellar medium will affect our data sets at this level and highlighted the need for multiple frequency observations in order to mitigate such effects.   

Timing residuals for PSR~J1909$-$3744 spanning 5.5\,yr from PPTA observations at an observing frequency $\sim$3\,GHz are shown in Figure~\ref{fg:1909}.  The rms timing residual is $\sim 100$\,ns showing that exquisite timing precision is achievable over such long time scales.  This data set is consistent with `white' noise and hence does not indicate any unmodelled physics (such as the presence of gravitational wave signals).  On even longer time scales, Verbiest et al. (2008) showed that PSR~J0437$-$4715 could be timed with an rms residual $\sim$200\,ns over ten years, but the timing residuals do show evidence of timing irregularities. Verbiest et al. (2009) followed up on this work by studying the timing residuals of 20 millisecond pulsars on time scales $> 10$\,yr. It was shown that the level of timing noise in most pulsars will not affect gravitational wave detection efforts on time scales of 5-10\,yr.  However, the timing residuals for many of the pulsars studied did indicate the presence of timing noise that could have resulted from irregular spin-down of the pulsars.  Verbiest et al. (2008) showed that even low-level timing noise in pulsar timing residuals may affect parameter estimation.  For instance, the published uncertainties on mass estimates that have been determined using pulsar timing need to be treated with caution if the timing residuals were affected by timing noise.

In a few cases, timing noise and glitches are clearly seen in the residuals for millisecond pulsars.  For instance, the timing residuals for PSR~B1937+21 have long been known (e.g. Kaspi, Taylor \& Ryba 1994) to be dominated by timing noise that is not caused by calibration procedures or the interstellar medium.    The first glitch in a millisecond pulsar was found in PSR~B1821$-$24 which is located in the globular cluster M28 (Cognard \& Backer 2004).  The glitch was shown to follow the main characteristics of glitches seen in the slower pulsars. 

\section{Conclusion}

Observations of pulsars over decadal timescales provides insight into the nature of pulsar interiors. The spin-rates and mass measurements that can be determined using pulsar timing constrain various models of the equation of state of the pulsar interior. Glitch events that occur in such long data spans provide a probe into the interaction between the interior and exterior of the star. Recent pulsar timing noise analyses has shown that glitch events cannot alone explain the timing irregularities observed and that pulsar timing noise may not be caused by effects in the stellar interior.

\section{Acknowledgements}

The Parkes radio telescope is part of the Australia Telescope, which is funded by the Commonwealth of Australia for operation as a National Facility managed by the Commonwealth Scientific and Industrial Research Organisation (CSIRO). GH is the recipient of an Australian Research Council QEII Fellowship (\#DP0878388).  Many people have been involved in timing pulsars from Jodrell Bank Observatory.  In particular I acknowledge Andrew Lyne, Michael Kramer and Christine Jordan.


\end{document}